\documentclass{svmult}

\usepackage{makeidx}         
\usepackage{graphicx}        
\usepackage{multicol}        
\usepackage[bottom]{footmisc}

\makeindex             
\usepackage{amssymb}
\usepackage{amsmath}
\usepackage[arrow,matrix,curve]{xy}

\newcommand{\alert}{\emph}
\newcommand{\rp}{\textup{\symbol{'051}}}
\newcommand{\lp}{\textup{\symbol{'050}}}
\newcommand{\wA}{\widehat{A}}
\newcommand{\fg}{\mathfrak{g}}
\newcommand{\cC}{\mathcal{C}}
\newcommand{\cH}{\mathcal{H}}
\newcommand{\cK}{\mathcal{K}}
\newcommand{\cL}{\mathcal{L}}
\newcommand{\cN}{\mathcal{N}}
\newcommand{\bC}{\mathbb{C}}
\newcommand{\bP}{\mathbb{P}}
\newcommand{\bT}{\mathbb{T}}
\newcommand{\bR}{\mathbb{R}}
\newcommand{\bZ}{\mathbb{Z}}

\newcommand{\co}{\colon\,}
\newcommand{\Ca}{$C^*$-algebra}
\newcommand{\Tr}{\operatorname{Tr}}
\newcommand{\Maps}{\operatorname{Maps}}
\newcommand{\vol}{\operatorname{vol}}
\newcommand{\Pic}{\operatorname{Pic}}
\newcommand{\Ind}{\operatorname{Ind}}
\newcommand{\Aut}{\operatorname{Aut}}
\newcommand{\Sq}{\operatorname{Sq}}
\newcommand{\pt}{\text{pt}}
\newcommand{\YM}{{\scriptstyle\text{YM}}}
\spnewtheorem{axioms}[theorem]{Axioms}{\bfseries}{\itshape}

\begin{document}

\title*{Dualities in Field Theories and the Role of $K$-Theory}
\titlerunning{Dualities and $K$-Theory}
\author{Jonathan Rosenberg\thanks{This research was partially
    supported by NSF grant 
  DMS-0805003. I would like to thank the organizers and
  participants of the Closing Meeting  on Perspectives in Deformation
  Quantization 
and Noncommutative Geometry in Kyoto, February 2011, as well as the staff of
the RIMS, for a very stimulating and well-run conference.}}
\institute{Department of Mathematics,
University of Maryland,
College Park, MD 20742--4015, USA, \texttt{jmr@math.umd.edu}}
%
%
\maketitle

%
\begin{abstract}
It is now known (or in some cases just believed)
that many quantum field theories exhibit \alert{dualities}, 
equivalences with the same or a different theory in which things
appear very different, but the overall physical implications 
are the same. We will discuss some of these dualities from the point
of view of a mathematician, focusing on ``charge 
conservation'' and the role played by $K$-theory and noncommutative
geometry. Some of the work described here is joint with Mathai
Varghese and Stefan Mendez-Diez; the last section is new.

\end{abstract}

\setcounter{minitocdepth}{1}
\dominitoc

\section{Overview with Some Classical Examples}
\label{part:intro}
\subsection{Structure of Physical Theories}
\label{sec:physthy}
Most physical theories describe 
\alert{fields}\index{field (in physics)}, e.g., the
\emph{gravitational field}, \emph{electric field}, \emph{magnetic field},
etc. Fields can be
\begin{itemize}
\item scalar-valued functions (\alert{scalars}),
\item sections of vector bundles (\alert{vectors}),
\item connections on principal bundles (\alert{special cases of
gauge fields})\index{field!gauge},
\item sections of spinor bundles (\alert{spinors}).
\end{itemize}

In classical physics, the fields satisfy a variational principle --- they
are critical points of the \alert{action}\index{action} 
$S$, which in turn is the integral of
a \emph{local} functional $\cL$ called the \alert{Lagrangian}. This is called
the \emph{principle of least action}\index{action!least}, 
and can be traced back to
Fermat's theory of optics in 1662. The Euler-Lagrange equations for
critical points of the action are the \alert{equations of motion}.

\begin{example}[Yang-Mills Theory]
Let $M$ be a $4$-dimensional Riemannian or Lorentzian manifold, say
compact. We fix a compact Lie group $G$ and a principal 
$G$-bundle over $M$. A Yang-Mills field is a connection $A$ on this
bundle. The ``field strength''\index{field strength} 
$F$ is the curvature, a $\fg$-valued $2$-form.
The action is $S=\int_M \Tr F\wedge *F$ (up to a constant involving
the coupling constant $g_{\YM}$ measuring the strength of the
interactions). Note that the metric on $M$ is 
needed to define the Hodge $*$-operator $F\mapsto *F$.
\end{example}
\begin{example}[General Relativity in Empty Space]
For convenience, we consider
the ``Wick rotation'' of the theory to Euclidean signature.
Let $M$ be a $4$-manifold, say compact.
A field is a Riemannian metric $g$ on $M$. The (Einstein-Hilbert) action is
$S=\int_M R\,\D\vol$, $R$ = scalar curvature. (Strictly speaking one
should insert a coupling constant in front, $\frac{c^4}{16\pi G}$,
where $G$ is Newton's gravitational constant and $c$ is the speed of
light, which we usually set equal to $1$ in suitable units.) The
associated field equation is Einstein's equation.
\end{example}

Unlike classical mechanics, quantum mechanics is not deterministic,
only probabilistic. The key property of quantum mechanics is the
\alert{Heisenberg 
uncertainty principle}\index{Heisenberg uncertainty principle}, 
that observable quantities
are represented by \emph{noncommuting operators} $A$
represented on a Hilbert space $\cH$. In the quantum
world, every particle has a wave-like aspect to it, and is represented
by a wave function\index{wave function}
$\psi$, a unit vector in $\cH$. The phase of $\psi$
is not directly observable, only its amplitude, or more precisely,
the \alert{state} $\varphi_\psi$ defined by $\psi$:
\[
\varphi_\psi(A) =\langle A\psi, \psi\rangle\, .
\]
But the phase is still
important since \emph{interference} depends on it.

The quantization of classical field theories is based on \emph{path
  integrals}.\index{path integral} 
The idea (not 100\% rigorous in this formulation) is
  that \emph{all fields contribute}, not just those that are critical
  points of the action (i.e., solutions of the classical field
  equations). Instead, one looks at the \alert{partition
    function}\index{partition function} 
\[
Z = \int \E^{\I S(\varphi)/\hbar}\,\D\varphi \text{ or } \int
  \E^{-S(\varphi)/\hbar}\,\D\varphi \, ,
\]
depending on whether one is working in Lorentz or Euclidean
signature. 
By the \alert{principle of stationary phase}\index{stationary phase},
only fields 
close to the classical solutions should contribute very much. 
\alert{Expectation values}\index{expectation value} 
of physical quantities are given by 
\[
\langle A\rangle = \left(\int A(\varphi)\,\E^{\I
  S(\varphi)/\hbar}\,\D\varphi\right)/ Z \, . 
\]

\subsection{Dualities}
\label{sec:duality}

A \alert{duality}\index{duality} 
is a transformation between different-looking
physical theories that, rather magically, have the same observable
physics. Often, such dualities are part of a discrete group, such
as $\bZ/2$ or $\bZ/4$ or $SL(2,\bZ)$.
\begin{example}[Electric-magnetic duality]
\label{ex:EMduality}\index{duality!electric-magnetic}
Let $E$ and $B$ be the electric and magnetic fields, respectively.
There is a symmetry of Maxwell's equations\index{Maxwell's equations}
in free space
\begin{equation}
\label{eq:Maxwell}
\begin{aligned}
\nabla\cdot E = 0, \quad & \quad \nabla\cdot B = 0,\\
\frac{\partial E}{\partial t} = c\,\nabla \times B,\quad &\quad
\frac{\partial B}{\partial t} = -c\,\nabla \times E,
\end{aligned}
\end{equation}
given by $E\mapsto -B$, $B\mapsto E$. This is a duality of order $4$.
\end{example}

\begin{example}[Configuration space-momentum space duality]
\label{ex:Fourierduality}\index{duality!Fourier}
Another example from standard quantum mechanics concerns the
quantum harmonic oscillator (say in one dimension).
For an object with mass $m$ and a restoring force with ``spring constant''
$k$, the Hamiltonian is
\begin{equation}
\label{eq:harmosc}
H = \frac{k}{2}x^2 + \frac{1}{2m}p^2\,,
\end{equation}
where $p$ is the momentum. In classical mechanics, $p = m\dot x$. But
in quantum mechanics (with $\hbar$ set to $1$),
\begin{equation}
\label{eq:Heisenberg}
[x,p]=\I\, .
\end{equation}
We obtain a duality of \eqref{eq:harmosc} and \eqref{eq:Heisenberg}
via $m\mapsto \frac{1}{k}$,
$k\mapsto \frac{1}{m}$, $x\mapsto p$, $p\mapsto -x$.
This is again a duality of order $4$, and is closely related to the
\alert{Fourier transform}\index{Fourier transform}.
\end{example}

\begin{itemize}
\item{A big puzzle in classical electricity and magnetism is that while
there are plenty of charged particles (electrons, etc.), no
magnetically charged particles (\alert{magnetic monopoles}) have ever
been observed, even though their existence would not contradict
Maxwell's equations.}\index{magnetic monopole}
\item{Another problem with classical electricity and magnetism is that
  it doesn't 
  explain why charges appear to be quantized, i.e., only occur in
  units that are integral multiples of the charge of the electron (or of
  the charges of [down-type] quarks).}\index{quantization!charge}
\end{itemize}

Dirac \cite{Dirac} proposed to solve both problems at
  once with a \alert{quantum} theory of electricity and magnetism that
  in modern terms we would call 
  a \alert{$U(1)$ gauge theory}.\index{Dirac monopole}

In Dirac's theory, we assume spacetime is a $4$-manifold $M$, say
$\bR^4\setminus \bR \cong \bR^2\times S^2$ (Minkowski space with the
time trajectory of one particle taken out). The
(magnetic) vector potential $(A^1,A^2,A^3)$ and electric potential $A^0=\phi$ of
classical electricity and magnetism 
are combined into a single entity $A$, a
(unitary) connection on a complex line bundle $L$ over $M$. Thus $\I A$
is \emph{locally} a real-valued
$1$-form, and $F=\I\mu\, \D A$, $\mu$ a constant, is a $2$-form
encoding both of the fields $E$ (via the $(0,j)$ components)
and $B$ (via the $(j,k)$ components, $0<j<k$). The Chern class $c_1(L)\in
H^2(M, \bZ)\cong \bZ$ is an invariant of the topology of the
situation. Of course, $F$ should really be $\I\mu$ times
the curvature of $A$, and Chern-Weil theory says that the de
Rham class $[F]$ is $2\pi \mu$ times the image of $c_1(L)$ in $H^2(M,
\bZ)\cong \bZ$. $L$ is associated to a principal $U(1)$-bundle $P\to  M$, and
Dirac identifies a section of this bundle with the phase of a wave
function of a charged particle in $M$.

In the above setup, if we integrate $F$ over the $S^2$ that links
the worldline we removed, we get $2\pi \mu c_1(L)$, and this is the
flux of the magnetic field through $S^2$. So the deleted worldline can
be identified with that of a \alert{magnetic monopole} of charge $g=\mu c_1(L)$
in suitable units. Suppose we consider the motion of a test charge of
electric charge $q$ around a closed loop $\gamma$ in $M$. In quantum
electricity and magnetism, by the 
Aharonov-Bohm effect\index{Aharonov-Bohm effect} \cite{AB},
the exterior derivative is replaced by the covariant derivative
(involving the vector potential $A$). So the phase change in
the wave function is basically the holonomy of $(P\to  M, A)$ around $\gamma$,
or (taking $\hbar=1$)
$\exp\left(q\mu\oint_\gamma A\right)$. Since $M$ is simply connected,
$\gamma$ bounds a disk $D$ and the integral is (by Stokes' Theorem) 
$\exp\left(-\I q\int_D F\right)$. 
Taking $D$ in turn to be the two hemispheres in $S^2$, we
get two answers which differ by a factor of 
\[
\exp\left(\I\, q\int_{S^2} F\right) = \E^{2\pi \I\,q\,\mu\,c_1(L)}.
\]
Since this must be $1$, we get \alert{Dirac's quantization
  condition}\index{quantization!Dirac condition}\index{Dirac
  quantization condition} 
$qg \in \bZ$.

The upshot of this analysis is that we expect both electrical and
magnetic charges to be quantized, but that 
\emph{the basic quanta of electrical and
magnetic charge should be inversely proportional in size}. In other
words, the smallness of the fundamental electrical charge means that
the charge of any magnetic monopole has to be large. In any event, we
expect the electrical and magnetic charges $(q,g)$ to take values in
an abelian \alert{charge group} $C$, in this case $\bZ^2$. It is also
reasonable to expect there to be particles, usually called
\alert{dyons}\index{dyon}, with both charges $q$ and $g$ non-zero.

Now think about the classical electric-magnetic duality  
(Example \ref{ex:EMduality}) that
switches $E$ and $B$. The \alert{Montonen-Olive
  Conjecture}\label{conj:MOconj}\index{Montonen-Olive Conjecture}
\cite{MO},  for which there is now some tantalizing evidence, is that
in a wide variety of cases this should extend to a duality of quantum
theories, which would necessarily give an isomorphism of charge groups
between a theory and its dual.

In Dirac's theory, the quantization of magnetic charge and of electrical
charge arise from different origins. The former is a purely
topological phenomenon; it comes from the fact that the Chern classes
live in \emph{integral} cohomology. Quantization of electrical charge
comes from the requirement that the \emph{action} (for the field
associated to a charged particle moving in the background
electromagnetic field of a monopole) be well-defined and
not multi-valued, so this can be viewed as a version of \alert{anomaly
cancellation}\index{anomaly cancellation}. 
However, since Maxwell's equations are invariant under
electro-magnetic duality, we can imagine an equivalent dual theory in which
electric charge is topological and magnetic charge is quantized to
achieve anomaly cancellation.

\subsection{A General Framework and the Role of $K$-Theory}
\label{sec:dualityframework}

Extrapolating from case above, we will be looking at the following
set-up:
\begin{enumerate}
\item We have a collection $\cC$ of ``physical theories'' on which a
  discrete \alert{duality group} $G$ operates by ``equivalences.''
  (More generally, $G$ might be replaced by a groupoid.)
\item Each theory in $\cC$ has an associated \alert{charge
  group}\index{charge group} 
  $C$. If $g\in G$ gives an equivalence between two theories in $\cC$,
  it must give an isomorphism between the associated charge groups. In
  particular, the stabilizer of a fixed theory operates by
  automorphisms on $C$.
\item In many cases, the charge groups arise as topological
  invariants. We have already seen how $\Pic X = H^2(X,\bZ)$
  arises. (The notation $\Pic X$ denotes the set of isomorphism
  classes of complex line bundles over $X$, which is a group under
  tensor product.)
  We will see how $K$-theory arises in some cases.
\end{enumerate}

Many of the most interesting examples of duality (and of topological
charge groups) arise in (supersymmetric) \alert{string
theories}\index{string theory}. 
These are quantum field theories based on the idea of
replacing point particles by \alert{strings} or $1$-manifolds (always
compact, but maybe with boundary --- contrary to mathematical usage,
physicists call these ``open strings''). For anomaly
cancellation\index{anomaly cancellation} 
reasons, the spacetime manifold has to be $10$-dimensional.
The worldsheet traced out by
a string in the spacetime $X$ is a compact $2$-manifold $\Sigma$ (again,
possibly with boundary), so we obtain fields that are maps $f\co\Sigma \to
X$, with the \alert{sigma-model
  action}\index{action!sigma-model}\index{sigma model} 
of the form
\begin{equation}
\label{eq:sigmodelaction}
\int_\Sigma \Vert\nabla f\Vert^2 + \int_\Sigma f^*(B) + (\text{terms
  involving other fields}).
\end{equation}
Here $B$ is a $2$-form on $X$ called the \alert{B-field}\index{B-field}
(\emph{not}
the magnetic field). The term $\int_\Sigma f^*(B)$ is called
the \alert{Wess-Zumino term}\index{Wess-Zumino term}.  The terms
involving the ``other fields'' depend on which of the five superstring
theories (type I, which allows unoriented strings, types IIA and IIB,
and the two types of heterotic theories)
one is dealing with.  They differ with regard to 
such issues as chirality and orientation conditions, and whether or not open
strings are allowed.

In string theories, boundary conditions (of Dirichlet or
  Neumann type) must be imposed on the open string states. These are
  given by \alert{D-branes}\index{D-brane} 
(D for ``Dirichlet''), submanifolds of the spacetime $X$ on which strings
are allowed to ``end.'' If we forget certain complications and look at
type II string theory, then $X$ is a $10$-dimensional spin manifold
and the stable D-branes are spin$^c$ submanifolds, of even dimension for type
IIB and of odd dimensional for type IIA. (At least in the absence of
twisting, $X$ is generally $\bR^4$ times a Calabi-Yau $3$-fold, and in
the type IIB case, the stable D-branes are complex submanifolds, whereas in
the IIA case, they are typically isotropic submanifolds for the
symplectic structure.)

There is another piece of
structure; each D-brane carries a 
\alert{Chan-Paton vector bundle}\index{Chan-Paton bundle}
that reflects a $U(N)$ gauge symmetry allowing for local exchanges between
coincident D-branes.

The D-branes carry \alert{charges} which are not just
  numbers but elements of the $K$-group $K(X)$ (in the type IIB
  theory), $K^{-1}(X)$ (in the type IIA theory), or $KO(X)$ (in the
  type I theory).

The idea that the D-brane charges should take values in \alert{$K$-theory}
comes from Minasian-Moore \cite{MR1606278} and Witten
\cite{MR1674715}, around 1997--1998, with further 
elaboration by other authors later. Motivation comes from several
sources: 
\begin{itemize}
\item compatibility with anomaly cancellation formulas;
\item better functoriality;
\item compatibility with analysis of decay of unstable branes;
\item compatibility with what is known about string duality.
\end{itemize}
We will not attempt to go through these arguments (which the reader
can find in \cite{MR1674715,MR1827946,MR1693636} and
\cite[\S6.2]{MR2285203}) but   will discuss some consequences.

For a D-brane $\xymatrix{W \ar@{^{(}->}[r]^\iota & X}$
with Chan-Paton 
bundle $E\to W$, the $K$-theory charge is $\iota_!([E])$, where $[E]$
is the class of $E$ in $K(W)$, and $\iota_!$ is the Gysin map\index{Gysin map}
in $K$-theory (defined using the spin$^c$ structures). While string
dualities do not have to preserve the diffeomorphism type, or even the
dimension, of D-branes, they do have to give rise to an isomorphism of
the $K$-groups in which the D-brane charges lie.

The most important kinds of string theory dualities
are \alert{T-duality}\index{T-duality}\index{duality!T-}, 
an outgrowth of classical Fourier
duality\index{duality!Fourier} 
(``T'' originally standing for ``target space''),
and \alert{S-duality}, an outgrowth of classical electro-magnetic
duality\index{duality!electric-magnetic}. 
The big difference between them is that T-duality preserves
coupling strength and changes geometry, whereas
S-duality\index{S-duality}\index{duality!S-}  
(``S'' standing for
``strong-weak'') interchanges strong and weak coupling and preserves
the geometry of spacetime, just as
electro-magnetic duality inverts the magnitude of charges.

Much of the interest of these dualities is that they are
\emph{non-perturbative}, in other words, don't depend on perturbation
expansions.  In some cases, a quantity which is difficult to compute
in one theory can be computed by passage to a dual theory in which
the quantity is easier to compute directly, or can be computed via
a perturbation expansion.

T-duality replaces tori (of a fixed dimension $k$) in the spacetime
manifold $X$ by their dual tori\index{dual torus} 
(quotients of the \emph{dual space} by
the \emph{dual lattice}) in the dual spacetime $X^\sharp$, inverting the radii. 
If $k$ is odd,
T-duality  interchanges the theories of types IIA and IIB, so one gets
an isomorphism $K(X) \cong K^{-1}(X^\sharp)$ or $K^{-1}(X) \cong
K(X^\sharp)$. S-duality interchanges type I string theory with the
\alert{$SO(32)$ heterotic} string theory, and also maps type IIB
string theory to itself.

In Sections \ref{part:topTdual} and \ref{part:SUdual}
I will discuss T-duality and S-duality in
more detail, and the way charge conservation in $K$-theory sheds more
light on them.
\section{Topological T-Duality}
\label{part:topTdual}
\subsection{The H-flux and Twisted $K$-Theory}
\label{sec:Hflux}

It's now time to correct a slight oversimplification in Section
\ref{part:intro}: the ``B-field'' in the sigma-model
action \eqref{eq:sigmodelaction} 
is not necessarily globally well-defined, though
its \emph{field strength}\index{field strength} 
$H=\D B$ does make sense globally. Properly
normalized, one can show that $H$ defines an integral de Rham class in
$H^3$. This can be refined to an actual class in $[H]\in H^3(X,\bZ)$.
Thus the Wess-Zumino term\index{Wess-Zumino term}
in the path integral should really be defined using
a \alert{gerbe}\index{gerbe}, 
for example a \alert{bundle gerbe} in the sense of
Murray \cite{MR1405064} with \emph{curving} $B$ and 
\emph{Dixmier-Douady class}\index{Dixmier-Douady class} $[H]$. We
usually refer to $H$ (or to the associated class $[H]\in H^3(X,\bZ)$) as
the \alert{H-flux}\index{H-flux}. (For an exposition of how gerbes can
be used to make sense of the Wess-Zumino term, see for example
\cite{MR1945806} or \cite[\S4.3]{MR2560910}.)

The association of $H$ with a Dixmier-Douady class is not
an accident, and indeed indicates a deeper connection with noncommutative
geometry. To set this up in the simplest way, choose
a stable \alert{continuous-trace algebra}\index{continuous-trace algebra} 
$A=CT(X,[H])$ with $\widehat A=X$ and
with Dixmier-Douady class\index{Dixmier-Douady class} 
$[H]$. Thus $A$ is the algebra of continuous
sections vanishing at $\infty$ of a bundle over $X$ with fibers $\cK$
(the compact operators on a separable $\infty$-dimensional Hilbert
space $\cH$) and structure group $\Aut \cK = PU(\cH)\simeq K(\bZ,2)$.

There are several possible definitions of \alert{twisted
$K$-theory}\index{twisted $K$-theory}
(see \cite{MR0282363,MR1018964,MR2172633,MR2307274,MR2513335}), but for our
purposes we can define it as $K^{-i}(M,[H]) = 
K_i(A)$ with $A=CT(M,[H])$ as above. Up to isomorphism, this only
depends on $X$ and the cohomology class $[H]\in H^3(X,\bZ)$.

In the presence of a topologically nontrivial H-flux, the
$K$-theoretic classification of D-brane\index{D-brane} 
charges has to be modified. A
D-brane $\xymatrix{W \ar@{^{(}->}[r]^\iota & X}$ in type II string
theory is no long a Spin$^c$ manifold; instead it is Spin$^c$ ``up to
a twist,'' according to the \alert{Freed-Witten anomaly cancellation
condition}\index{anomaly cancellation} \cite{MR1797580}
$W_3(W) = \iota^*([H])$. Here $W_3$ is the canonical integral lift of
the third Stiefel-Whitney class, which is the obstruction to a
Spin$^c$ structure. Accordingly, the D-brane charge
will live in the \alert{twisted $K$-group} $K(X, [H])$ (in type IIB)
or in $K^{-1}(X, [H])$ (in type IIA).  Accordingly, if we have a
T-duality between string theories on 
$(X,H)$ and $(X^\sharp,H^\sharp)$, \alert{conservation of
charge} (for D-branes) requires an isomorphism of twisted $K$-groups
of $(X,[H])$ and $(X^\sharp,[H^\sharp])$, with no degree shift if we dualize with
respect to even-degree tori, and with a degree shift if we dualize with
respect to odd-degree tori.

One might wonder what happened to the $K$-groups of
opposite parity, viz., $K^{-1}(X, [H])$ (in type IIB) and $K(X, [H])$
(in type IIA). These still have a physical significance in terms
of \alert{Ramond-Ramond fields}\index{field!Ramond-Ramond} 
\cite{MR1827946}, so want these to match up under T-duality also.

\subsection{Topological T-Duality and the Bunke-Schick Construction}
\label{sec:topTduality}

\alert{Topological
  T-duality}\index{duality!T-!topological}\index{T-duality!topological}  
focuses on the topological aspects of
T-duality. The first example of this phenomenon was studied by
Alvarez, Alvarez-Gaum\'e, Barb\'on, and Lozano in 1993 \cite{MR1265457}, and
generalized 10 years later by Bouwknegt, Evslin, and Mathai
\cite{MR2080959,MR2116165}.  
Let's start with the simplest nontrivial example of a circle fibration,
where $X=S^3$, identified with $SU(2)$, $T$ is a maximal torus.
Then $T$ acts freely on $X$ (say by right translation) and the
quotient $X/T$ is $\bC\bP^1\cong S^2$, with quotient map
$p\co X\to S^2$ the \alert{Hopf fibration}. Assume for simplicity
that the $B$-field vanishes. We have $X=S^3$ fibering over
$Z=X/T=S^2$. Think of $Z$ as the union of the two hemispheres
$Z^\pm \cong D^2$ intersecting in the equator $Z^0 \cong S^1$. The
fibration is trivial over each hemisphere, so 
we have $p^{-1}(Z^\pm)\cong D^2\times S^1$,
with $p^{-1}(Z^0)\cong S^1\times S^1$. So the T-dual also
looks like the union of two copies of $D^2\times S^1$, joined along
$S^1\times S^1$.

However, we have to be careful about the \alert{clutching} that
identifies the two copies of $S^1\times S^1$. In the original Hopf
fibration, the clutching function $S^1\to S^1$ winds once around, with
the result that the fundamental group $\bZ$ of the fiber $T$ dies in
the total space $X$. But T-duality is supposed to interchange
``winding'' and ``momentum'' quantum numbers. So the T-dual $X^\sharp$ has no
winding and is just $S^2\times S^1$, while the winding of the original
clutching function shows up in the $H$-flux of the dual.

In fact, following Buscher's method \cite{MR901282}
for dualizing a sigma-model, we find that the $B$-field $B^\sharp$
on the dual side is different on the two
copies of $D^2\times S^1$; they differ by a closed $2$-form, and so
$H^\sharp=\D B^\sharp$, the H-flux of the dual, is nontrivial in de
Rham cohomology (for simplicity of notation we
delete the brackets from now on) but well defined.

Let's check the principle of $K$-theory matching in the case we've
been considering, $X=S^3$ fibered by the Hopf fibration over $Z=S^2$.
The $H$-flux on $X$ is trivial, so D-brane charges lie in
$K^*(S^3)$, with no twisting. And $K^0(S^3)\cong K^1(S^3)\cong \bZ$.

On the T-dual side, we expect to find $X^\sharp=S^2\times S^1$, also
fibered over $S^2$, but simply by projection onto the first factor. If
the $H$-flux on $X$ were trivial, D-brane changes would lie in
$K^0(S^2\times S^1)$ and $K^1(S^2\times S^1)$, both of which are
isomorphic to $\bZ^2$, which is \alert{too big}.

On the other hand, we can compute $K^*(S^2\times S^1,
H^\sharp)$ for the class $H^\sharp$ which is $k$ times a generator
of $H^3\cong\bZ$, using the Atiyah-Hirzebruch Spectral Sequence. The
differential is 
\[
H^0(S^2\times S^1) \xrightarrow{k} H^3(S^2\times S^1),
\]
so when $k=1$, $K^*(S^2\times S^1, H^\sharp)\cong K^*(S^3)\cong \bZ$ for both
$*=0$ and $*=1$.

\subsubsection{Axiomatics for $n=1$}

This discussion suggests we should try to develop an axiomatic
treatment of the \alert{topological} aspects of T-duality (for circle
bundles).  Note that 
we are ignoring many things, such as the underlying metric on
spacetime and the auxiliary fields. Here is a first attempt.

\begin{axioms}
\label{sec:TDualAxioms}
\leavevmode
\begin{enumerate}
\item We have a suitable class of spacetimes $X$ each equipped with a
  principal $S^1$-bundle $X\to Z$. {\lp}$X$ might be required to be a
  smooth connected manifold.{\rp}
\item For each $X$, we assume we are free to choose any $H$-flux $H\in
  H^3(X,\bZ)$. 
\item There is an involution {\lp}map of period $2${\rp} $(X,H) \mapsto
  (X^\sharp, H^\sharp)$ keeping the base $Z$ fixed.
\item $K^*(X, H) \cong  K^{*+1}(X^\sharp, H^\sharp)$.
\end{enumerate}
\end{axioms}

\subsubsection{The Bunke-Schick Construction}
\label{sec:BunkeSchick}
Bunke and Schick \cite{MR2130624}
suggested constructing a theory satisfying these
axioms by means of a \alert{universal example}. It is known that
(for reasonable spaces $X$, say CW complexes) all principal
$S^1$-bundles $X\to Z$ come by \alert{pull-back} from a diagram
\[
\xymatrix{X \ar[d] \ar@{.>}[r] & ES^1\simeq * \ar[d] \\
Z \ar@{.>}[r] & BS^1 \simeq K(\bZ, 2)}
\]
Here the map $\xymatrix{ Z \ar@{.>}[r] & K(\bZ,2)}$ is unique up to
homotopy, and pulls the canonical class in $H^2(K(\bZ,2),\bZ)$ back to
$c_1$ of the bundle.

Similarly, every class $H\in H^3(X,\bZ)$ comes by pull-back from a 
canonical class via a map $\xymatrix{X \ar@{.>}[r] & K(\bZ,3)}$ unique up to
homotopy.

\begin{theorem}[{Bunke-Schick \cite{MR2130624}}]
\label{thm:BunkeSchick}
There is a classifying space $R$, unique up to homotopy equivalence,
with a fibration
\begin{equation}
\label{eq:BSPost}
\xymatrix{K(\bZ,3) \ar[r] & R \ar[d] \\
& \,K(\bZ, 2)\times K(\bZ,2),}
\end{equation}
and any $(X,H)\to Z$ as in the axioms comes by a pull-back
\vspace{-2pt}
\[
\xymatrix{X \ar[d] \ar@{.>}[r] &E \ar[d]^p \\
Z \ar@{.>}[r] & \,R,}
\]
with the horizontal maps unique up to homotopy and $H$ pulled back
from a canonical class $h\in H^3(E,\bZ)$.
\end{theorem}

\begin{theorem}[{Bunke-Schick \cite{MR2130624}}]
\label{thm:BunkeSchick1}
Furthermore, the $k$-invariant of the Postnikov tower
\eqref{eq:BSPost}
characterizing $R$ is the cup-product in
\[
H^4(K(\bZ, 2)\times K(\bZ,2),\bZ)
\]
of the two canonical classes in $H^2$. The space $E$ in the fibration
\[
\xymatrix{S^1 \ar[r] & E \ar[d]^p \\
& R}
\]
has the homotopy type of $K(\bZ, 3)\times K(\bZ,2)$.
\end{theorem}

\begin{corollary}
If $(X\xrightarrow{p}Z, H)$ and $(X^\sharp\xrightarrow{p^\sharp}Z,
H^\sharp)$ are a T-dual pair of circle 
bundles over a base space $Z$, then the bundles and fluxes are related
by the formula
\[
p_!(H) = [p^\sharp], \quad (p^\sharp)_!(H^\sharp) = [p].
\]
Here $[p]$, $[p^\sharp]$ are the Euler classes of the bundles, and
$p_!$, $(p^\sharp)_!$ are the ``integration over the fiber'' maps in
the Gysin sequences. Furthermore, there is a pullback diagram of
circle bundles
\[
\xymatrix{Y \ar[r]^{(p^\sharp)^*(p)} \ar[d]_{p^*(p^\sharp)}& X \ar[d]^p \\
X^\sharp \ar[r]^{p^\sharp}& Z.}
\]
in which $H$ and $H^\sharp$ pull back to the same class on $Y$.
\end{corollary}

\subsubsection{The Case $n>1$}
\label{ref:higherdimTdual}

We now want to generalize T-duality to the case of spacetimes $X$
``compactified on a higher-dimensional torus,'' or in other words,
equipped with a principal $\bT^n$-bundle $p\co X\to Z$. In the
simplest case, $X=Z\times \bT^n=Z\times\overbrace{S^1\times\cdots
S^1}^n$.  We can then perform a string of $n$ T-dualities, one
circle factor at a time. A single T-duality interchanges type IIA and
type IIB string theories, so this $n$-dimensional T-duality
``preserves type'' when $n$ is even and switches it when $n$ is
odd. In terms of our Axioms \ref{sec:TDualAxioms}
for topological T-duality, we would therefore expect an
isomorphism $K^*(X, H) \cong  K^{*}(X^\sharp, H^\sharp)$ when $n$ is
even and $K^*(X, H) \cong  K^{*+1}(X^\sharp, H^\sharp)$ when $n$ is
odd. 

In the higher-dimensional case, a new problem
presents itself: 
\alert{it is no longer clear that the T-dual should be unique}.
In fact, if we perform a string of $n$ T-dualities, one
circle factor at a time, it is not clear that the result should be
independent of the order in which these operations are done.
Furthermore, a higher-dimensional torus does not split as a product in
only one way, so in principle there can be a lot of non-uniqueness.

The way out of this difficulty has therefore been to try
to organize the information in terms of a 
\alert{T-duality group}\index{T-duality group}, a
discrete group of T-duality isomorphisms potentially involving a large
number of spacetimes and $H$-fluxes. We can think of this group as
operating on some big metaspace of possible spacetimes.

Another difficulty is that there are some spacetimes with
$H$-flux that 
would appear to have no higher-dimensional T-duals at all, at least in
the sense we have defined them so far, e.g., $X=T^3$,
viewed as a principal $\bT^3$-bundle over a point, with
$H$ the generator of $H^3(X,\bZ)\cong \bZ$. 

\subsection{The Use of Noncommutative Geometry}
\label{sec:NCTdual}
Here is the strategy of the Mathai-Rosenberg approach
\cite{MR2116734,MR2327179,MR2222224}.  Start with a principal
$\bT^n$-bundle $p\co X\to Z$ and an ``$H$-flux'' $H\in H^3(X,\bZ)$. We
assume that $H$ is trivial when restricted to each $\bT^n$-fiber of
$p$. This of course is no restriction if $n=2$, but it rules out cases
with no T-dual in any sense.

We want to lift the free action of $\bT^n$ on $X$ to an action on the
continuous-trace algebra $A=CT(X,H)$. Usually there is no hope to get
such a lifting for $\bT^n$ itself, so we go to the universal covering
group $\bR^n$. If $\bR^n$ acts on $A$ so that the induced action on
$\wA$ is trivial on $\bZ^n$ and factors to the given action of
$\bT^n=\bR^n/\bZ^n$ on $\wA$, then we can take the crossed product
$A\rtimes \bR^n$ and use Connes' Thom Isomorphism Theorem to get an
isomorphism between $K^{-*-n}(X,H)=K_{*+n}(A)$ and   $K_*(A\rtimes
\bR^n)$.

Under favorable circumstances, we can hope that the
  crossed product 
$A\rtimes \bR^n$  will again be a continuous-trace algebra
$CT(X^\sharp, H^\sharp)$, with $p^\sharp\co X^\sharp\to Z$ a new
principal $\bT^n$-bundle and with $H^\sharp\in H^3(X^\sharp, \bZ)$. If
we then act on $CT(X^\sharp, H^\sharp)$ with the dual action of
$\widehat{\bR}^n$, then by Takai
Duality and stability, we come back to where we started. So we
have a topological T-duality between $(X,H)$ and
$(X^\sharp,H^\sharp)$. Furthermore, we have an isomorphism
\[
K^{*+n}(X,H)\cong K^*(X^\sharp,H^\sharp),
\]
as required for matching of D-brane charges under T-duality in Axioms
\ref{sec:TDualAxioms} (as modified for $n\ge 1$).

\alert{Now what about the problems we identified before, about
  potential non-uniqueness of the T-dual and ``missing'' T-duals?}
These can be explained either by non-uniqueness of the lift to an
action of $\bR^n$ on $A=CT(X,H)$, or else by failure of the crossed product
to be a continuous-trace algebra.

\subsubsection{A Crucial Example}
\label{ex:missingTdual}
Let's now examine what happens when we try to carry out this program
in one of our ``problem cases,'' $n=2$, $Z=S^1$, $X=T^3$ (a trivial
$\bT^2$-bundle over $S^1$), and $H$ the usual generator of $H^3(T^3)$.
First we show that there is an action of $\bR^2$ on $CT(X,H)$
compatible with the free action of $\bT^2$ on $X$ with quotient
$S^1$. We will need the notion of an \emph{induced action}.  We start
with an action $\alpha$ of $\bZ^2$ on $C(S^1,\cK)$ which is 
trivial on the spectrum. This is given by a map $\bZ^2\to C(S^1,\Aut
\cK) = C(S^1, PU(L^2(\bT)))$ sending the two generators of $\bZ^2$ to
the maps
\[
\begin{aligned}
w&\mapsto \text{multiplication by } z,\\
w&\mapsto \text{translation by } w,
\end{aligned}
\]
where $w$ is the coordinate on $S^1$ and $z$ is the coordinate on $\bT$.
(Of course $S^1$ and $\bT$ are homeomorphic, but we use different
letters in order to distinguish them, since they play slightly
different roles. These two unitaries commute in $PU$, not in $U$.)

Now form $A=\Ind_{\bZ^2}^{\bR^2}C(S^1,\cK)$. This is a {\Ca} with
$\bR^2$-action $\Ind \alpha$ whose spectrum (as an $\bR^2$-space) is
$\Ind_{\bZ^2}^{\bR^2}S^1 = S^1 \times \bT^2 = X$. We can see that $A
\cong CT(X,H)$ via ``inducing
in stages''. Let $B=\Ind_\bZ^\bR C(S^1, \cK(L^2(\bT)))$ be the
result of inducing over the first copy of $\bR$. 
It's clear that $B\cong C(S^1\times \bT,
\cK)$. We still have another action of $\bZ$ on $B$
coming from the second generator of $\bZ^2$, and
$A=\Ind_\bZ^\bR B$. The action of
$\bZ$ on $B$ is by means of a map $\sigma\co
S^1\times \bT \to PU(L^2(\bT))=K(\bZ,2)$, whose value
at $(w,z)$ is the product of multiplication by $z$ with
translation by $w$.  Thus $A$ is a CT-algebra with Dixmier-Douady invariant
$[\sigma]\times c=H$, where $[\sigma]\in 
H^2(S^1\times \bT,\bZ)$ is the homotopy class of $\sigma$ and $c$ is
the usual generator of $H^1(S^1,\bZ)$. 

Now that we have an action of $\bR^2$ on $A=CT(X,H)$
  inducing the free 
$\bT^2$-action on the spectrum $X$, we can compute the crossed product
to see what the associated ``T-dual'' is.
Since $A=\Ind_{\bZ^2}^{\bR^2}C(S^1,\cK)$, we can use the
Green Imprimitivity Theorem to see that
\[
A\rtimes_{\Ind \alpha} \bR^2 \cong \Bigl(C(S^1,\cK)\rtimes_{\alpha}
\bZ^2 \Bigr)\otimes \cK .
\]

Recall that \alert{$A_\theta$} is the universal {\Ca} generated by
unitaries $U$ and $V$ with $UV=e^{2\pi i \theta} VU$. So if we look at
the definition of $\alpha$, we see that $A\rtimes_{\Ind \alpha} \bR^2$
is the algebra of sections of a bundle of algebras over $S^1$, whose
fiber over $e^{2\pi i \theta}$ is $A_\theta\otimes
\cK$. Alternatively, it is Morita equivalent to $C^*(\Gamma)$, where
$\Gamma$ is the \alert{discrete Heisenberg group} of strictly
upper-triangular $3\times 3$ integral matrices.

Put another way, we could argue that we've shown that $C^*(\Gamma)$ is
a \alert{noncommutative 
T-dual}\index{duality!T-!noncommutative}\index{T-duality!noncommutative} 
to $(T^3,H)$, both viewed as fibering
over $S^1$. So we have an explanation for the missing T-dual:
\alert{we couldn't find it just in the world of topology alone because
  it's noncommutative}. We will want to see how widely this phenomenon
occurs, and also will want to resolve the question of nonuniqueness of
T-duals when $n>1$.

Further analysis of this example leads to the following classification
theorem:
\begin{theorem}[{Mathai-Rosenberg \cite{MR2116734}}]
\label{thm:MR1}
Let $\bT^2$ act freely on $X=T^3$ with quotient $Z=S^1$. Consider the set
of all actions of $\bR^2$ on algebras $CT(X,H)$ inducing this action
on $X$, with $H$ allowed to vary over $H^3(X,\bZ)\cong \bZ$. Then the
set of exterior equivalence classes of such actions is parametrized
by $\Maps(Z,\bT)$. The winding number of a map $f\co Z\cong\bT\to\bT$ can be
identified with the Dixmier-Douady invariant $H$. All these actions
are given by the construction above, with $f$ as the ``Mackey
obstruction map.''
\end{theorem}

Consider a general $\bT^2$-bundle $X\xrightarrow{p} Z$.
We have an edge homomorphism 
\[
p_!\co H^3(X,\bZ)\to E_\infty^{1,2} \subseteq H^1(Z,H^2(\bT^2,\bZ))
= H^1(Z,\bZ)
\]
which turns out to play a major role.
\begin{theorem}[{Mathai-Rosenberg\cite{MR2116734}}]
\label{thm:MR3}
Let $p\co X\to Z$ be a principal $\bT^2$-bundle as above, $H\in
H^3(X,\bZ)$. Then we can always find a ``generalized T-dual'' by
lifting the action of $\bT^2$ on $X$ to an action of $\bR^2$ on
$CT(X,H)$ and forming the crossed product. When $p_!H=0$, we can
always do this in such a way as to get a crossed product of the form
$CT(X^\sharp,H^\sharp)$, where $(X^\sharp,H^\sharp)$ is a classical
T-dual {\lp}e.g., as found though the purely topological
theory{\rp}. When $p_!H\ne 0$, the crossed product $CT(X,H)\rtimes
\bR^2$ is \alert{never locally stably commutative} and should be viewed as a
\alert{noncommutative T-dual}.
\end{theorem}

\subsection{Current Directions in Topological T-Duality}
\label{sec:currentTduality}
Here we just summarize some of the current trends in topological
T-duality\index{duality!T-!topological}\index{T-duality!topological}:
\begin{enumerate}
\item the above approach with actions of $\bR^n$ on continuous-trace
algebras.  For $n\ge 2$, the lift of even a free action of $\bT^n$ on
$X$ to an action of $\bR^n$ on $CT(X,H)$ is usually not essentially
unique, and a more detailed study of non-uniqueness is required.
One would also like to extend the study of topological T-duality to
cases where the action of $\bT^n$ has isotropy, as progress on the
famous ``SYZ conjecture'' \cite{MR1429831} will require study of torus
bundles with some degeneration. These issues have been studied in
\cite{MR2222224}, \cite{MR2246781}, and \cite{MR2369414}, for example.
\item the homotopy-theoretic approach of Bunke-Schick, extended to
the higher-dimensional case. This has been studied by
Bunke-Rumpf-Schick \cite{MR2287642}, by
Mathai-Rosenberg \cite{MR2222224}, and by Schneider \cite{Schneider}.
\item a fancier approach using duality of sheaves over the
  Grothendieck site of (suitable) topological spaces
(Bunke-Schick-Spitzweck-Thom \cite{MR2482327}). 
\item a generalization of the noncommutative geometry approach using
  groupoids (Daenzer   \cite{MR2491618}). 
\item algebraic analogues, in the world of complex manifolds, schemes,
  etc., using Mukai duality with gerbes (Ben-Basset,
Block, Pantev \cite{MR2309993} and Block and Daenzer \cite{MR2608194}).
\item work of Bouwknegt and Pande \cite{MR2672468} relating the
  noncommutative geometry 
  approach to Hull's notion of T-folds \cite{MR2180623}, which are
  certain nongeometric backgrounds well-known in string theory.  
\item an approach of Bouwknegt and Mathai using duality for loop group
  bundles \cite{MR2539268}.
\end{enumerate}

As one can see, this is a very active subject going off in many
different directions, and it would take a much longer survey to go
into these matters in detail.

\section{Problems Presented by S-Duality and Other Dualities}
\label{part:SUdual}

\subsection{Type I/Type IIA Duality on $T^4$/K3}
\label{sec:typeItypeII}
In this subsection I want to describe some joint work with Stefan
Mendez-Diez \cite{MendezRos}. As we mentioned before, there is
believed to be an S-duality\index{S-duality}\index{duality!S-} relating 
type I string theory to one of the heterotic string theories.  There
are also various other dualities relating these two theories to type
IIA theory.  Putting these together, we expect a (non-perturbative) duality
between \alert{type I string theory on $T^4\times \bR^6$} and \alert{type
IIA string theory on $K3\times \bR^6$}, at least at certain points in the
moduli space. (Here $K3$ denotes a 
K3 surface\index{K3 surface}, a simply connected closed
complex surface with trivial canonical bundle. The name K3
stands for ``Kummer, K\"ahler, Kodaira.'' As a
manifold, it has Betti numbers $1$, $0$, $22$, $0$, $1$, and signature
$-16$.) This duality is discussed in detail in
\cite{MR2010972}.
How can we reconcile this with the principle that brane
charges in type I should take their values in $KO$, while brane
charges in type IIA should take their values in $K^{-1}$?

On the face of it, this appears ridiculous:
$KO(T^4\times \bR^6) = KO^{-6}(T^4)$ has lots of $2$-torsion, while
$K^*(K3)$ is all torsion-free and concentrated in \alert{even} degree.

One side is easy compute. Recall that for any space $X$,
\[
KO^{-j}(X\times S^1)\cong KO^{-j}(X)\oplus KO^{-j-1}(X).
\]
Iterating, we get
\[
\begin{aligned}
KO^{-6}(T^4)&\cong KO^{-6} \oplus 4KO^{-7} \oplus 6 KO^{-8} \oplus
4KO^{-9} \oplus KO^{-10}\\
&\cong \bZ^6 \oplus (\bZ/2)^4 \oplus (\bZ/2) \cong \bZ^6 \oplus (\bZ/2)^5.
\end{aligned}
\]

The way we deal with the opposite side of the duality is to recall
that a K3 surface can be obtained by blowing up the point singularities
in $T^4/G$, where $G=\bZ/2$ acting by multiplication by $-1$ on
$\bR^4/\bZ^4$. This action is semi-free with $16$ fixed points, the
points with all four coordinates equal to $0$ or $\frac12$ mod $\bZ$.
If fact one way of deriving the (type I on $T^4$) $\leftrightarrow$
(type IIA on $K3$) duality explicitly uses the orbifold $T^4/G$.

But what group should orbifold brane charges live in? 
$K^*(T^4/G)$ is not quite right, as this ignores the orbifold structure. One
solution that has been proposed is $K^*_G(T^4)$, which Mendez-Diez and I
computed. However, as we'll see, there appears to be a better
candidate.

Let $M$ be the result of removing a
$G$-invariant open ball around each $G$-fixed point
in $T^4$. This is a compact \alert{manifold with boundary} on which $G$
acts \alert{freely}; let $N=M/G$. We get a  K3 surface\index{K3 surface} 
back from $N$ by
gluing in $16$ copies of the unit disk bundle of the tangent bundle of
$S^2$ (known to physicists as the Eguchi-Hanson 
space\index{Eguchi-Hanson space}), one along each
$\bR\bP^3$ boundary component in $\partial N$. 
\begin{theorem}[{\cite{MendezRos}}]
\[
H^i(N,\partial N)\cong H_{4-i}(N) \cong 
\left\{\begin{array}[pos]{ll} 0, & i=0\\ 
\bZ^{15}, & i=1\\
\bZ^6, & i=2\\
(\bZ/2)^5, & i=3\\
\bZ, & i=4\\ 
0, & \text{otherwise}.\\
\end{array} \right.\]
\end{theorem}

Recall $N$ is the manifold with boundary obtained from $T^4/G$ by
removing an open cone neighborhood of each singular point.
\begin{theorem}[{\cite{MendezRos}}]
$K^0(N, \partial N) \cong K_0(N)  \cong \bZ^7$ and 
$K^{-1}(N, \partial N)  \cong  K_1(N) \cong \bZ^{15} \oplus(\bZ/2)^5$. 
\end{theorem}
Note that the reduced $K$-theory of $(T^4/G)\mod (\text{singular
  points})$ is the same as $K^*(N, \partial N)$. Note
the resemblance of $K^{-1}(N, \partial N)$
to $KO^{-6}(T^4) \cong \bZ^6 \oplus (\bZ/2)^5$. While they are not the
same, the calculation suggests that the brane charges in type I string
theory on $T^4\times \bR^6$ do indeed show up some way in type IIA string
theory on the orbifold limit of $K3$.

Again let $G=\bZ/2$. Equivariant $K$-theory $K_G^*$ is a module over
the \alert{representation ring} $R=R(G)=\bZ[t]/(t^2-1)$. This ring
has two important prime ideals, $I=(t-1)$ and $J=(t+1)$. We have
$R/I\cong R/J\cong \bZ$, $I\cdot J = 0$, $I+J=(I,2)=(J,2)$,
$R/(I+J)=\bZ/2$.
\begin{theorem}[{\cite{MendezRos}}]
$K_G^0(\bT^4)\cong R^8\oplus(R/J)^8$, and $K_G^{-1}(\bT^4)=0$. 
Also,\hbox{\vrule height8pt depth8pt width0pt}
$K^0_G(M, \partial M)\cong (R/I)^7$, $K^{-1}_G(M,\partial M)\cong
(R/I)^{10}\oplus (R/2I)^5$.
\end{theorem}

Note that the equivariant $K$-theory calculation is a refinement of
the ordinary $K$-theory calculation (since $G$ acts freely on $M$ and
$\partial M$ with quotients $N$ and $\partial N$, so that $K^*_G(M)$
and $K^*_G(\partial M)$ are the same 
as $K^*(N)$ and $K^*(\partial N)$ \emph{as abelian groups}, though
with the addition of more structure). While we don't immediately need
the extra structure, it may prove useful later in matching brane
charges from $KO(T^4\times \bR^6)$ on specific classes of branes.

\subsection{Other Cases of Type I/Type II Charge Matching}
More generally, one could ask if there are circumstances where
understanding of $K$-theory leads us to expect the possibility of a
string duality between type I string theory on a spacetime $Y$ and
type II string theory on a spacetime $Y'$. For definiteness, we will
assume we are dealing with type IIB on $Y'$. (This is no great loss of
generality since as we have
seen in Section \ref{part:topTdual}, types IIA and IIB are
related via T-duality.)  Matching of stable D-brane\index{D-brane} 
charges then leads us to look for an isomorphism of the form
\[
KO^*(Y) \cong K^*(Y').
\]
In general, such isomorphisms are quite rare, in part because of
$2$-torsion in $KO^{-1}$ and $KO^{-2}$, and in part because
$KO$-theory is usually $8$-periodic rather than $2$-periodic.

But there is one notable exception: one knows \cite[p.\ 206]{MR1324104} 
that 
\[
KO\wedge (S^0 \cup_\eta e^2)\simeq K,
\]
where $S^0 \cup_\eta e^2$ is the stable cell complex
obtained by attaching a stable $2$-cell via the stable $1$-stem
$\eta$. This is stably the same (up to a degree shift)
as $\bC\bP^2$, since the attaching map
$S^3\to S^2\cong \bC\bP^1$ for the top cell of $\bC\bP^2$ is the Hopf
map, whose stable homotopy class is $\eta$. Thus one might expect a
duality between type I string theory on
$X^6\times \bigl( \bC\bP^2 \smallsetminus \{\pt\}\bigr)$ and type IIB
string theory on $X^6\times \bR^4$. We plan to look for evidence for this.

\subsection{The AdS/CFT Correspondence}

The \alert{AdS/CFT correspondence}\index{AdS/CFT correspondence}
or \alert{holographic duality}
is a conjectured physical duality,
proposed by Juan Maldacena \cite{MR1633016}, of a 
different sort, relating IIB string theory on a $10$-dimensional
spacetime manifold to a gauge theory on another space. In the original
version of this duality, the string theory lives on $AdS^5\times S^5$,
and the gauge theory is $\cN=4$ super-Yang-Mills theory on Minkowski
space $\bR^{1,3}$.  Other versions involve slightly different spaces
and gauge theories. A good survey may be found in \cite{MR1743597}.
Notation:
\begin{itemize}
\item $\cN$ is the standard notation for the \alert{supersymmetry
multiplicity}. In other words, $\cN=4$ means
there are $4$ sets of supercharges, and
there is a $U(4)$ \alert{R-symmetry}\index{R-symmetry} group acting on them.
\item $AdS^5$, $5$-dimensional \emph{anti-de Sitter
  space}\index{anti-de Sitter space} is (up to coverings) the
  homogeneous space  $SO(4,2)/SO(4,1)$. Topologically, 
this homogeneous space is $\bR^4\times S^1$. It's better to pass to the
universal cover $\bR^5$, however, so that time isn't periodic.
\end{itemize}

\subsubsection{Nature of the Correspondence}

We have already explained that D-branes\index{D-brane} carry Chan-Paton
bundles. In 
type IIB string theory, a collection of $N$ coincident D3 branes have
$3+1=4$ dimensions and carry a $U(N)$ gauge theory living on the
Chan-Paton bundle\index{Chan-Paton bundle}. 
This gauge theory is the holographic dual of the
string theory, and the number $N$ can be recovered as the flux of the
Ramond-Ramond (RR) field\index{field!Ramond-Ramond} strength $5$-form
$G_5$ through an $S^5$ linking the D3 brane
\cite[equation (3.7)]{MR1743597}. The rotation group $SO(6)$
of $\bR^5$ is identified (up to coverings)
with the $SU(4)_R$ symmetry group of the $\cN=4$ gauge theory.

The AdS/CFT correspondence 
looks like holography\index{holography} in that physics in 
the bulk of AdS space is described by a theory of one less dimension
``on the boundary.'' This can be explained by the famous
Beckenstein-Hawking bound for the entropy of a black hole in terms of the area
of its boundary, which in turn forces quantum gravity theories to obey
a \alert{holographic principle}.

Recall that the Montonen-Olive
Conjecture\index{Montonen-Olive Conjecture} (Section
\ref{conj:MOconj}) asserts that classical electro-magnetic
duality should 
extend to an exact symmetry of certain quantum field
theories. $4$-dimensional super-Yang-Mills\index{super-Yang-Mills theory}
(SYM) with $\cN=4$
supersymmetry is believed to be a case for which this conjecture
applies. The Lagrangian involves the usual Yang-Mills term
\[
\frac{-1}{4g^2_{\YM}}\int \Tr(F\wedge *F)
\]
and the \alert{theta angle}\index{theta angle}
term (related to the Pontrjagin number or
\alert{instanton number}\index{instanton number})
\[
\frac{\theta}{32\pi^2}\int \Tr(F\wedge F).
\]
We combine these by introducing the \alert{tau parameter}
\[
\tau = \frac{4\pi i}{g^2_{\YM}} + \frac{\theta}{2\pi}.
\]

The tau parameter measures the relative size of ``magnetic'' and
``electric charges.'' Dyons\index{dyon} 
in SYM have charges $(m,n)$ living in the
group $\bZ^2$; the associated \emph{complex charge} is $q+ig =
q_0(m+n\tau)$. As in the theory of the Dirac monopole\index{Dirac
monopole}, quantization of magnetic charge is related to integrality
of characteristic classes in topology, i.e., to the fact
that the Pontrjagin number must be an integer.

The \alert{electro-magnetic duality group}\index{duality!electric-magnetic}
$SL(2,\bZ)$ acts on
$\tau$ by linear fractional transformations. More precisely, it is
generated by two transformations: $T\co \tau\mapsto \tau+1$, which
just increases the $\theta$-angle by $2\pi$, and has no effect on
magnetic charges,
and by $S\co \tau\mapsto -\frac{1}{\tau}$, which effectively
interchanges electric and magnetic charge. By the Montonen-Olive
Conjecture\index{Montonen-Olive Conjecture}
\cite{MO}, the same group $SL(2,\bZ)$ should operate on type IIB
string theory in a similar way, and $\theta$ should correspond in the
string theory to the expectation value of the RR scalar
field\index{field!Ramond-Ramond}  $\chi$. (See for example
\cite{MR1321523,MR1394827,MR1377168,MR1743597}.)

\subsubsection{Puzzles About Charge Groups}
An important constraint on variants of the AdS/CFT correspondence
should come from the action of the $SL(2,\bZ)$ S-duality group on the
various charges. For example, this group is expected to act on the
pair $(H,G_3)$ of type IIB string theory field strengths 
in $H^3(X,\bZ)\times H^3(X,\bZ)$ by linear fractional
transformations. Here $G_3$ denotes the RR $3$-form field strength, or more
precisely, its cohomology class. But now we have some puzzles: 
\begin{itemize}
\item The classes of RR fields are really supposed
to live in $K^{-1}$, not cohomology, whereas the NS class $[H]$ is
really expected to live in ordinary cohomology. (Fortunately, since the first 
differential in the Atiyah-Hirzebruch spectral sequence is $\Sq^3$,
there is no difference when it comes to classes in $H^3$, except when
$H^3$ has $2$-torsion. See  \cite{MR1975994,MR2250274} for related
discussions.) 
\item Since the S-duality group mixes the NS-NS and RR sectors, it
is not clear how it should act on D-brane and RR field charges.
\item It's also not so clear what conditions to impose at infinity
when spacetime is not compact. For example, it would appear that the
H-flux and RR fields do not have to have compact support, so perhaps
$K$-theory with compact support is not the right home for the RR field
charges. This point seems unclear in the literature.
\end{itemize}

\begin{example}
Let's look again at the example of type IIB string
theory on $AdS^5\times S^5$, compared with $\cN=4$ SYM on $4$-space.
How do the $K$-theoretic charge groups match up? Our spacetime is
topologically $X= \bR^5\times S^5$, where $\bR^5$ is the universal cover of
$AdS^5$. We think of $\bR^5$ more exactly as $\bR^4\times \bR_+$, so
that $\bR^4\times \{0\}$, Minkowski space, is ``at the boundary.'' The
RR field charges should live in $K^{-1}(X)$, according to
\cite{MR1827946}, but we see this requires
clarification: the RR field strength $G_5$ should represent the number $N$ in
$H^5(S^5)$ (since as we mentioned, $N$ is computed by pairing the
class of $G_5$ with the fundamental class of $S^5$), so we need to use
\alert{homotopy theoretic} $K$-theory $K_h$ 
here instead of $K$-theory with compact support, which we've
implicitly been using before. Indeed, note that $K^{-1}(X) \cong K^{-1}(\bR^5)
\otimes K^0(S^5)\cong H^0(S^5)$, while $K^{-1}_h(X) \cong K_h^{0}(\bR^5)
\otimes K^{-1}(S^5)\cong H^5(S^5)$, which is what we want.

Now what about the D-brane\index{D-brane} charge group for the string
theory? This should be $\bZ \cong K^0(X) \cong K^0(\bR^4 \times
Y)\cong K^0(\bR^4)\otimes K^0(Y)$,
where $Y$ is the D5-brane $\bR\times S^5$, which has
$K^0(Y)\cong \bZ$. Note that this is naturally isomorphic to
$K^0(\bR^4) = \widetilde K^0(S^4)$, which is where the \alert{instanton
number}\index{instanton number} 
lives in the dual gauge theory. But what charge group on $X$
corresponds to the 
group of electric and magnetic charges in the gauge theory?
(This should be a group isomorphic to $\bZ^2$ containing the group
$\bZ$ classifying the instanton number.)
\end{example}

It is believed that the string/gauge correspondence should apply much
more generally, to many type IIB string theories on spaces other than
$AdS^5 \times S^5$, and to gauge theories with less supersymmetry than
the $\cN=4$ theory that we've been considering.  Analysis of the
relevant charge groups on both 
the string and gauge sides of the correspondence should give us a
guide as to what to expect. Study of these
constraints is still in a very early stage.

\bibliographystyle{amsplain}
\bibliography{Rosenberg-Kyoto11}

\printindex
\end{document}